\documentstyle[prl,multicol,aps]{revtex}
\input{epsf}

\newcommand{\bq}{\begin{equation}}
\newcommand{\ee}{\end{equation}}
\newcommand{\fr}[2]{\frac{#1}{#2}}
\newcommand{\eps}{\varepsilon}

\begin{document}
\draft

\title{ Stretched Exponential Decay of a Quasiparticle in a
Quantum Dot}

\author{P.G.Silvestrov}

\address{
Instituut-Lorentz, Universiteit Leiden, P.O. Box 9506, 2300
RA Leiden, The Netherlands,\\
and Budker Institute of Nuclear Physics, 630090
Novosibirsk, Russia
}

\maketitle
\date{today}

\begin{abstract}

The decay of a quasiparticle in an isolated 
quantum dot is considered. At relatively small time
the probability to find the system in the initial state decays
exponentially: $P(t)\sim \exp(-\Gamma t)$, in accordance with the
golden rule. However, the contributions to $P(t)$ accounting
for the discreteness of final three-particle states, five-particle 
states, etc. decay much slower being $\sim (\Delta_3/\Gamma)^n 
\exp(-\Gamma t/(2n+1))$ for $2n+1$ final particles.  Here $\Delta_3 
\ll \Gamma$ is the level spacing for three-particle states available 
via the direct decay. These corrections are dominant at large enough 
time and slow down the decay to become $\ln (P)\sim -\sqrt{t}$ 
asymptotically. $P(t)$ fluctuates strongly in this regime and 
the analytical formula for the distribution $W(P)$ is found.

\end{abstract}


\begin{multicols}{2}

The problem of the decay of a quasiparticle in closed
mesoscopic systems has attracted a permanent interest last
years~\cite{SImry,Sivan,Blanter,AGKL,Mirlin,sil,Avishai,Rupp,Carlo,Folk}.
The very existence of the (irreversible) decay becomes nontrivial
in the case of a discrete spectrum. 
In a many particle system  one should in addition understand the role
of the whole hierarchy of the discreteness, ranging from the 
single-particle levels (level spacing $\Delta$) to the exponentially
dense spectrum of complicated many-particle excitations.
It was first realized that the decay into two particles and 
one hole, relevant for an infinite geometry, does not apply
for the low excitation energies in the finite system.
Namely, for a diffusive quantum dot, the coupling to the
three-particle states describes the broadening of the quasiparticle 
only for excitation energy $\eps\gg \sqrt{g}\Delta$, where 
$g\gg 1$ is the dimensionless conductance~\cite{Sivan,Blanter,AGKL}.
Utilization of the ideas of delocalization in the Fock space
for the quasiparticle decay in Ref.~\cite{AGKL} has
stimulated further active research in this 
field~\cite{Mirlin,sil,Avishai,Rupp,Carlo}.
However, the main interest was concentrated around the limited 
range of excitation energies
$\sqrt{g/\ln(g)}<\eps/\Delta<\sqrt{g}$.
In spite of the beauty of the applied theoretical approaches 
experimentally it may be difficult to look for the effects 
sensitive to 
the logarithm of large parameter (see however~\cite{Folk}).
Therefore the goal of this paper is to consider the effect of 
the discreteness of the spectrum for much higher energy, where the
usual three-particle decay is expected to 
determine the quasiparticle lifetime.
As we will see, even for $\eps\gg \sqrt{g}\Delta$
the decay in the confined geometry at large enough time
is strongly modified and slowed down.

The simple fermionic Hamiltonian models the effect of electron 
interaction in closed quantum dot~\cite{AGKL}
(in the nuclear physics an analogous Hamiltonian was introduced
already a long time ago \cite{Wong,Bohigas}) 
\bq\label{Ham}
H=\sum \eps_i c_i^+c_i +\sum V_{ijkl}c_i^+c_j^+c_kc_l \ .
\ee
The single-particle energies $\eps_i$ are randomly distributed
around $\epsilon_F=0$ with mean level spacing $\Delta$.
The depth of the Fermi sea is always much larger than the
excitation energy. The Gaussian random two-particle
interaction has zero mean and
variance given by~\cite{Blanter,AGKL}
\bq\label{g}
\overline{ V^2} ={\Delta^2}/{g^2} .
\ee
The conductance $g\gg 1$ measured in units of $e^2/h$ is a large 
parameter in our problem. 
We are interested in the stability of 
simple single-particle excitations above the ground state of 
$H$. Due to the weak interaction the low
lying excited states are almost unperturbed. However, 
the level spacing for three-particle
states (two particles and one hole) accessible for the direct decay 
decreases while increasing the excitation energy $\eps$~\cite{sil}
\bq\label{delta3}
\Delta_3={4\Delta^3}/{\eps^2}
\ee
and at
$\eps >\sqrt{g}\Delta$
the matrix element of the interaction (\ref{g})
exceeds $\Delta_3$. 
Now the Fermi golden rule is applied to find
the width of the quasiparticle
\bq\label{Gamma}
\Gamma=2\pi\fr{\overline{|V|^2}}{\Delta_3}
=\fr{\pi\eps^2}{2g^2\Delta} \ .
\ee
The natural quantity characterising the time evolution of the
single-particle state is the {\it return
probability} $P(t)$, which is the probability to find a system
in the initial state $|0\rangle$ after time $t$
\bq\label{PP}
P
=|\langle 0|e^{-iHt}|0\rangle|^2=
\int e^{i(\eps'-\eps)t}G_{00}(\eps)G^*_{00}(\eps')
\fr{d\eps' d\eps}{4\pi^2} .
\ee
Here we have introduced $G_{00}(\eps)$, the diagonal matrix 
element of the single-particle Greens function of the 
Hamiltonian~(\ref{Ham}).

The authors of Ref.~\cite{AGKL} have used the Cayley tree model in
order to describe the quasi-particle disintegration into
many-particle states. 
One site of the lattice (tree) is associated with the single-particle 
state which is connected with many three-particle states (sites). 
Each of the three-particle states is connected with a number of 
five-particle states, and so on.
On the Cayley tree the {\it exact}
Greens function is given by the simple formula~\cite{Cayley}
\bq\label{Gree}
G_{00}(\eps)=\fr{1}{\eps_\lambda-\eps_0-{\displaystyle \sum_n\fr{V_n^2}
{\eps_\lambda-E_n-{\displaystyle \sum_m\fr{V_{nm}^2}
{\eps_\lambda-E_{m}-...}}}}} ,
\ee
where $\eps_\lambda=\eps+i\lambda $ 
with some small $\lambda$. Each new denominator of the
continued fraction corresponds to a mixing with the next generation (a
set of states in the Fock space having one more excited particle
and one more hole). 
$E_n=\eps_{n_1}+\eps_{n_2}+\eps_{n_3}$,
$E_m=\eps_{m_1}+\eps_{m_2}+\eps_{m_3}+\eps_{m_4}+\eps_{m_5}$ 
and so on. $V_n$ is the matrix element 
connecting the
single-particle and three particle states and $V_{nm}$ is the 
matrix element
connecting the $n$-th three-particle and $m$-th five-particle
states. All matrix elements satisfy Eq.~(\ref{g}). 
Not all effects of the two-particle interaction are
taken into account by the Cayley tree model~\cite{sil}. 
Nevertheless, one may
associate any term of the perturbative expansion (in $V$)
of Eq.~(\ref{Gree}) with a certain class of true diagrams of
many-body perturbation theory. The diagrams notpresent
in Eq.~(\ref{Gree}) should be taken into account
separately, but they will not change
the results of this paper.
In particular, the interaction
of the initial particle with the three-particle states
is completely taken into account by the Cayley tree.
Replacing the first sum in the denominator 
in Eq.~(\ref{Gree}) by
the integral over averages, one gets
\bq\label{G00}
\overline{G_{00}(\eps)}={\cal{G}}_0(\eps)=
(\eps-\eps_0+i\Gamma/2)^{-1} 
\ ,
\ee
where the width
\bq\label{Gam}
\Gamma=2\int\fr{\overline{V^2}}{\eps_{\lambda}-x+i\Gamma_n/2}
\fr{dx}{\Delta_3}= 2\pi \fr{\overline{V^2}}{\Delta_3}
\ee
essentially does not depend on the secondary width $\Gamma_n$.
The pole of ${\cal{G}}_0$ leads to the exponential decay
of the return probability $P(t)\sim\exp(-\Gamma t)$ and it is
hard to change this result while considering separately
$\overline{G(\eps)}$ and $\overline{G^*(\eps')}$. The
slow tail of $P(t)$ at large $t$  may come only from 
terms in the correlation function
$\overline{G_{00}(\eps)G^*_{00}(\eps')}$ that are non-analytic
in $\eps-\eps'$.

Let at first stage forget about 
the decay of
secondary states in Eq.~(\ref{Gree}) (i.e.  we put  
$V_{nm}\equiv 0$).
Now the expansion of
the Greens function (\ref{Gree}) around the averaged value
(Eq.~(\ref{G00})) gives
\bq\label{expan}
G_{00}=
{\cal{G}}_0+
{\cal{G}}_0^2
\left\{
\sum_n\fr{V_n^2}{\eps_{\lambda}-E_n} -
\int\fr{\overline{V^2}}{\eps_{\lambda}-x}\fr{dx}{\Delta_3}
\right\} 
.
\ee
Again $\eps_{\lambda}=\eps+i{\lambda}$.
Only the correlation of the second term $\{...\}$ in
$G_{00}(\eps)$ with its analog in $G^*_{00}(\eps')$ may lead to
the correction 
singular in $\eps-\eps'$. It is convenient to replace the first
sum in Eq.~(\ref{expan}) also by an integral via $\sum
F(E_n)\rightarrow \int\sum \delta (x-E_n)F(x)dx$. Now the only
nontrivial part in the calculation of the correlation function
is the averaging~\cite{note} 
\begin{eqnarray}\label{avav}
&&\overline{\Biggl(\sum_n V_n^2\delta(x-E_n)
-\fr{\overline{V^2}}{\Delta_3} 
\Biggr) 
\Biggl(\sum_m V_m^2\delta(y-E_m) -\fr{\overline{V^2}}{\Delta_3}
\Biggr)}
\nonumber\\
&& \ \ \ \ \ \ \ \ \ \ \ \ \ \ \ \ \ \ \ \ \ \ \ \ \ 
 =\fr{\overline{V^4}}{\Delta_3} \delta(x-y)\ . 
\end{eqnarray}
We suppose that $\eps_n$ are independent random variables (Poisson
statistics). In fact the problem of the decay of a simple state into
the quasi-continuum of discrete states (just like our decay into
three particles) has already been considered by several
authors~\cite{Mello,Sokolov,Seligman}. In particular, the case
of constant interaction and Wigner-Dyson statistics for secondary
states is considered in~\cite{Mello}.

With the use of Eqs.~(\ref{expan},\ref{avav}) one finds 
(${\cal{G}}'_0\equiv{\cal{G}}_0(\eps')$)
\begin{eqnarray}\label{expan1}
 \overline{G_{00}(\eps)G_{00}^*(\eps')}=
{\cal{G}}_0{\cal{G}}_0'^* 
+({\cal{G}}_0{\cal{G}}_0'^*)^2
 \fr{2\pi i}{\eps'-\eps-2i\lambda}\fr{\overline{V^4}}{\Delta_3}
\ ,
\end{eqnarray}
which after substitution into Eq.~(\ref{PP}) gives (for
Gaussian random interaction $\overline{V^4}=3\Delta^4/2g^4$)
\bq\label{Pav}
\overline{P(t)}
=e^{-\Gamma t}+ \fr{3\Delta_3}{2\pi\Gamma}
\left( \fr{\Gamma}{\Gamma-2\lambda}\right)^3
e^{-2\lambda t}
.
\ee
Here $2\lambda$ works effectively as the constant intrinsic
width of the three-particle excited states. After the initial 
state spreads over the $\sim\Gamma/\Delta_3$ directly connected states,
the usual Breit-Wigner decay saturates. It is important for this 
result that the interaction matrix elements and(or) the three-particle 
interlevel spacings essentially fluctuate. Otherwise the coherent
decay may proceed up to much larger times ($\Delta_3^{-1}$) and
even may become reversible.

The many particle origin of the final states is in fact 
ignored in Eq.~(\ref{Pav}).
In the real problem the energy and width of the ``intermediate''
final state are given by the sum of energies and widths of
the constituent particles $E_n=\eps_1+\eps_2+\eps_3\approx \eps$ and 
$\Gamma_n=\sum\Gamma_i\sim \sum\eps_i^2$
(\ref{Gamma}). In order to take into account this energy
dependence of the secondary width one should return to 
Eq.~(\ref{delta3}) and consider the three-particle density of
states as the sum over individual particle energies. This means
that in Eq.~(\ref{Pav}) the following replacement should be made 
\begin{eqnarray}
\fr{e^{-2\lambda t}}{\Delta_3} \ \rightarrow \
\int
&&
\exp\left\{ -\sum_{i=1}^3\Gamma_i t\right\}
\delta(\eps-\sum\eps_i)\prod\fr{d\eps_i}{\Delta}
\\
&&=
\fr{2}{\sqrt{3}}\fr{g^2}{\Delta^2t}e^{-\Gamma t/3} \ .
\nonumber 
\end{eqnarray}
The return probability now takes the form
\bq\label{15}
\overline{P}=e^{-\Gamma t}+\fr{27\sqrt{3}}{4}\fr{\Delta_3}{\Gamma}
\fr{1}{\Gamma t}e^{-\Gamma t/3} + ... \ .
\ee
Here the states from the second generation are allowed to decay
and their widths are distributed within $\Gamma/3 <\Gamma_n
<\Gamma$. However, the fluctuations of the widths, and even
more important, correlations of the fluctuations, are still not taken
into account. The second term in the r.h.s. of Eq.~(\ref{15})
is not small necessarily and for $\Gamma t>(3/2)L$ even
becomes dominant. Here we introduced a special notation
for the large logarithm $L=\ln (\Gamma/\Delta_3)$.
However even at these times the prefactor $1/\Gamma t$
in the return probability is only logarithmically small.
We will neglect the contributions of this type
in the calculation of further corrections to the return probability.

In the way similar to the derivation of Eq.~(\ref{expan}) we 
may extract from
the Greens function the term responsible for fluctuations in the
coupling of three-particle to five-particle states
\begin{eqnarray}\label{expan2}
&&G_{00}= 
{\cal{G}}_0+...+
{\cal{G}}_0^2
\sum_n{\cal{G}}_n^2{V_n^2}
\\ 
&&\times\int \left( \sum_m V_{nm}^2\delta(x-E_{m}) -
\fr{\overline{V^2}}{\Delta_{n}}\right)\fr{dx}{\eps_{\lambda}-x}+... \ . 
\nonumber
\end{eqnarray}
Here $\Delta_{n}\sim \Delta_3$ is the interval between the
five-particle states into which the given three-particle state
$n$ may decay.  

The connected average of the five-particle term in 
Eq.~(\ref{expan2}) with the same contribution from $G_{00}^*(\eps')$
leads to a pole at $\eps'-\eps=i\Gamma/5$ in complete analogy 
with Eq.~(\ref{expan1}). Here again $\Gamma/5$ is the smallest 
total width for the five particles with given total energy
$\sum\eps_i=\eps$. The straightforward estimation of the 
five-particle contribution to the return probability gives
\bq\label{P5}
\overline{P_5}
\sim \left( {\Delta_3}/{\Gamma}\right)^2 e^{-\Gamma t/5} 
\ee
Neither the overall numerical constant nor the ``weak'' 
$\sim 1/\Gamma t$ prefactor are shown here.
Non-Cayley-tree contributions to $\overline{P_5}$ coming
from the corrections to $G_{00}$ not included in the continued
fraction Eq.~(\ref{Gree}) are $\Gamma/\Delta$ times smaller 
and decay 
also like $e^{-\Gamma t/5}$ (or faster) at large time $t$.

Generalisation of Eq.~(\ref{P5}) for arbitrary
number of final particles gives
\bq\label{Pn}
\overline{P_{2n+1}}\sim \left( {\Delta_3}/{\Gamma}\right)^n
e^{-\Gamma t/(2n+1)} \ . 
\ee
Adding one more particle-hole pair in the final state costs a
small factor ${\Delta_3}/{\Gamma}$, but the time evolution
of this contribution is governed by the smallest joint width for
$2n+1$ particles with fixed total energy, which is
$\Gamma/(2n+1)$. 
The total return probability is given by the sum over all
many particle contributions~(\ref{Pn}) (starting from $n=0$). 
However, at any given moment of time one of the $\overline{P_{2n+1}}$
dominates and all others may be neglected.
Figure~1 shows the function $\ln P(t)$ and illustrates how
it is formed by different contributions  from Eq.~(\ref{Pn}).
The two consecutive values $\overline{P_{2n+1}}$ coincides and the 
strength of decay is changed at $\Gamma t= (2n-1)(2n+1)L/2$.
For large $\Gamma t$ one may also replace the piecewise linear
$\ln P$ by single smooth function (see Figure) such that
\bq\label{asym}
\overline{P_{asym}}\sim e^{ -\sqrt{2L
\Gamma t \ }} 
=\exp\left\{ -\sqrt{\ln\left(
\fr{\eps}{\sqrt{g}\Delta}\right)\fr{4\pi\eps^2}{g^2\Delta} t \
}\right\} . 
\ee
This formula should be compared with the usual decay $\exp
(-\Gamma t)=\exp \{ -\pi\eps^2t/2g^2\Delta\}$.  Eq.~(\ref{asym})
is the main result of this paper. Since the true many particle
density of states is $\delta^{-1}\sim
\exp(-2\pi\sqrt{\eps/6\Delta})$ this decay may formally proceed
until time $t\sim (g^2/\eps) 1/\ln(\eps/\sqrt{g}\Delta)$.

Within the interval ${3}L/2<\Gamma t< {15}L/2$
the second term in Eq.~(\ref{15}) dominates. Consider the
distribution of $P(t)$ in this region. The
direct generalisation of the calculation described by 
Eqs.~(\ref{PP},\ref{expan},\ref{expan1}) gives
\bq\label{moment}
\overline{P(t)^n}=n!\overline{P(t)}^n \ .
\ee
Here $n!$ simply accounts for the number of ways how $n$ Greens 
functions $G_{00}(\eps_i)$ may be contracted with
$n$ $G_{00}^*(\eps'_j)$. The distribution function corresponding
to the eq.~(\ref{moment}) is evidently 
\bq\label{distrib}
W(P)={1}/{\overline{P}}\exp\left\{ -{P}/{\overline{P}} \right\} \ .
\ee
Moreover, at any time only one contribution~(\ref{Pn}),
corresponding to fixed number of final particles, dominates
in $P(t)$. Therefore, Eq.~(\ref{moment}) and consequently 
Eq.~(\ref{distrib}) are also valid
for any large time $\Gamma t> 3L/2$. The strong fluctuations of 
$P(t)$ in the asymptotics~(\ref{distrib}) are in contrast with the
usual decay of the return probability at $\Gamma t< 3L/2$,
where one has ${\rm var} P(t)\ll P(t)^2$.

It is seen from Eq.~(\ref{distrib}) that the stretched 
asymptotic decay of a quasiparticle in a finite system
described by Eq.~(\ref{asym}) does not require any
special realization of the quantum dot. The interesting 
asymptotic effects due to rare fluctuations of disorder 
were considered more than a decade ago in Ref.~\cite{Dyhne}.
(See also the recent discussion of the decay in case of 
strong interaction~\cite{Flambaum}.)

All the above results only correspond to the case of relatively 
high excitation energy. Below $\eps\sim\sqrt{g}\Delta$ the
width expected from the golden rule become smaller then the 
level spacing for available three-particle states 
$\Gamma<\Delta_3$~(\ref{Gamma}).
Still at $\sqrt{g/\ln g}\ll \eps/\Delta \ll 
\sqrt{g}$ quasiparticles are unstable~\cite{AGKL}. Not
much is known about the time dependence of the return
probability $P(t)$ in this region. The decay now is not 
exponential at any time. One may convert the analysis of 
Ref.~\cite{sil} in order to get the time dependent series 
\bq\label{lowe}
\overline{P_{\Gamma<\Delta_3}}=1-\fr{\eps^2}{g\Delta^2}\sum_{n=0}c_n
\left( \fr{\eps^2}{g\Delta^2} \ln \left( \fr{\Delta}{g}t
\right)\right)^n  ,
\ee
where the $c_n$ are some (unknown) numerical
coefficients. 
Unfortunately, Eq.~(\ref{lowe}) is 
valid only for $1-\overline{P}\ll 1$. This result may be used 
for an estimation of the time after which the decay of
the return probability starts, but it does not teach us
about the functional form of $P(t)$.

To conclude, in this paper we have considered the decay of a
single-particle excitation in a finite fermionic system with
random two-particle interaction. Our results may
be interpreted as follows. At short time $t$ the decay into
two-particles and one hole proceeds with the Breit-Wigner width
$\Gamma$, in accordance with the golden rule. At a
certain time the probabilities to find the system in the initial
or any of the {\it discrete} final three-particle states became of the same
order of magnitude.  After that the further decay (exponential
on average) is governed by smaller width $\Gamma/3$,
which is the minimal width of the corresponding three-particle
state. At these times the probability $P(t)$ to find the initial 
quasiparticle unperturbed fluctuates strongly, but we
were able to find the distribution function
$W(P)$ describing these fluctuations~(Eq. (\ref{distrib})). 
When the three-particle states become
completely mixed with the five-particle states, the
five-particle width come into play leading to
$P(t)\sim\exp(-\Gamma
t/5)$, and so on. This scenario with many stage decay takes
place for the random interaction and an essentially irregular
single-particle spectrum, which suppresses the interference
between different decaying channels.
At sufficiently large time this piece-wise
linear exponential decay may be described by the smoothed
formula (\ref{asym}) with $\ln(P)\sim -\sqrt{t}$. All these
results are valid for high enough excitation energy 
($\eps\gg\sqrt{g}\Delta$).
The behaviour of $P(t)$ below the golden rule threshold, at
$\eps <\sqrt{g}\Delta$, 
requires a further investigation.

Fruitful discussions with Y.~Imry have stimulated 
the author to work on this problem.
The hospitality of the Weizmann Institute of Science (Israel)
where this project was started
is greatly acknowledged. 
I am thankful to Y.~Aharonov, C.~W.~J.~Beenakker, D.~V.~Savin 
and V.~V.~Sokolov 
for useful correspondence. 
This work was supported by the Dutch Science Foundation NWO/FOM.
The work in Israel was supported by
the Albert Einstein Minerva Center for Theoretical Physics and by
grants from the German-Israeli Foundation (GIF) and the Israel 
Science Foundation, Jerusalem.

\vspace{-.4cm}
\begin{figure}[t]
\epsfxsize=8.6cm
\epsffile{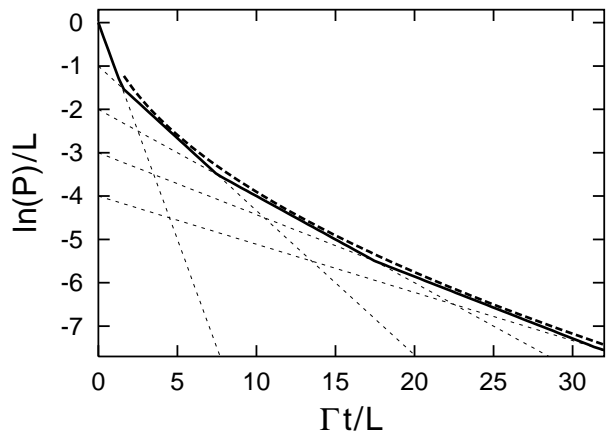}
\vglue 0.2cm
\medskip
\caption{The logarithm of the return probability
$P$ as a function of $\Gamma t$, both measured in units
of large logarithm $L=\ln(\Gamma/\Delta_3)$ (thick polygonal line).
The thin dashed lines are the individual contributions due to the
three-particle decay ($-\Gamma t$) and due to the correlations
in the $2n+1$-particle final state ($-\Gamma t/(2n+1)$). The thick
dashed line is the smoothed asymptotics 
$P\sim \exp\{-\sqrt{2L\Gamma t}\}$.}
\end{figure}

\end{multicols}


\begin{thebibliography}{99}
\bibitem{note} To perform the averaging it is convenient to
write each multiplier here as a sum of two terms
\[
\sum(V_n^2-\overline{V^2})\delta(x-\eps_n) + \overline{V^2}
[(\sum \delta(x-\eps_n)-\Delta_3^{-1}] \ ,
\]
one responsible for the fluctuations of the potential $V_n$ and
another for the positions of the levels $\eps_n$.
\end{thebibliography}
\end{document}